# Effect of anodization conditions on the synthesis of TiO$_2$ nanopores

Subhasish Chatterjee[1,2], Miriam Ginzberg[1], and Bonnie Gersten[1]
[1]Chemistry, Queens College, 65-30 Kissena Blvd., Flushing, NY, 11367
[2]Chemistry, The Graduate Center of CUNY, New York, NY, 10016

## ABSTRACT

Nanoporous structures play a promising role in the development of nanomechanical, nanoelectrical and biosensing devices. Recently, the voltage-driven transport of polynucleic acids (e.g. DNA, RNA) across synthetic inorganic nanopores has been investigated as a step towards the development of nanoscale biosensing devices. In addition, nanopores are expected to be utilized as chemical and gas sensors. TiO$_2$ is a semiconductor material which can have a wide range of applications in nanopore-based sensors. In this study, TiO$_2$ nanopores were prepared by electrochemical anodization. Titanium was used as the anode, while platinum was used as the cathode in an electrochemical cell filled with a hydrofluoric acid electrolyte solution. During the preparation process, titanium was converted to its oxide form. Nanostructures were synthesized under varying physical conditions, including HF concentrations of 0.5-10% and anodization times of 5-30 minutes. The resulting nanopore structures were characterized by scanning electron microscopy (SEM). The results show that the dimensions and morphology of the nanopores can be controlled by alteration of the anodization conditions.

## INTRODUCTION

Nanoporous materials have drawn significant interest, because of their versatile applications in fields such as optics [1], electronics [2], catalysis [3] and biosensing [4]. Various nanostructures, including nanopores and nanotubes, have been fabricated to manufacture nanoscale devices for chemical and biosensing processes [5]. Titanium dioxide (TiO$_2$) has emerged as an important oxide material in the development of nanopore based sensors [4]. The high refractive index (n=2.4) and semiconductive properties of TiO$_2$ are of great importance in photocatalytic and gas sensing processes [6, 7]. Since anodic oxidation of titanium metal can produce nano-architectured porous materials [8], the anodization process is a way to design large arrays of nanopores with variable size, shape and morphology.

Recently, various in-vitro experiments have been conducted to describe the voltage driven transport of single stranded (ss) DNA through synthetic nanoporous structures [9,10] and protein, suspended in a lipid layer [11]. The investigation of the voltage driven translocation of polynucleic acids (e.g. DNA, RNA) through nanopore structures has given rise to the idea of developing novel nanopore based biosensing devices and DNA sequencing techniques [11]. Since inorganic nanostructures display chemical, mechanical and thermal rigidity in diverse experimental conditions, their nanoporous environment can be adjusted [12]. Accordingly, the solid-state nanopores made of oxide materials have a great advantage over biological nanopores in the development of biosensors [4, 12].

As a part of a continuing endeavor to utilize $TiO_2$ nanopores in the development of a nucleic acid sensor, the effects of anodization conditions in the formation of $TiO_2$ nanoporous arrays were examined. As the morphology and dimensions of the nanopores significantly influence the sensitivity and selectivity of nanoscale devices, the ability to control these parameters by modifying the anodization conditions is imperative for sensor development. Previous studies have described the effect of anodizing voltage and pH on $TiO_2$ nanopore dimensions and morphology [13, 14]. This study aimed toward an understanding of the role of anodization time and electrolyte concentration in the synthesis of a large array of titanium oxide nanopores, at constant voltage.

## EXPERIMENTAL PROCEDURE

Titania nanopores were prepared by anodizing titanium foil in electrolyte solutions containing hydrofluoric acid (HF). Titanium foils with a thickness of 0.25 mm and 99.7% purity were purchased from Sigma-Aldrich. The foil was cleaned by sonication in a solution of methanol, acetone, and isopropanol for 20 minutes. A Teflon electrochemical cell, which maintained single-sided contact between the electrodes and the electrolyte solution and held the electrodes at a distance of 2 mm apart, was used for the experiment. The metal foil was placed as the anode in the electrochemical cell, with a platinum cathode. HF solutions of different concentrations (0.5%, 1%, 5%, and 10%) were prepared in ultrapure water. Approximately 2 mL of electrolyte solution was used for each anodization process, and 2.27 $cm^2$ of the titanium surface was exposed to the electrolyte solution. Using a power supply (VWR AccuPower), a constant potential of 20 V was applied to run the process, with varying anodization times (5, 10, 20, and 30 minutes). To protect the oxide layer from dissolution, the $TiO_2$ sample was rapidly removed from the cell after anodization. Afterwards, the sample was rinsed with deionized water and dried with nitrogen gas. A scanning electron microscope (SEM, Hitachi S-2600) was used to characterize the nanopore structures.

**RESULTS AND DISCUSSION**

Titanium oxide nanopores were successfully grown by anodic oxidation. Two principle anodization parameters, anodization time and concentration of HF electrolyte solutions, were regulated during the synthesis.

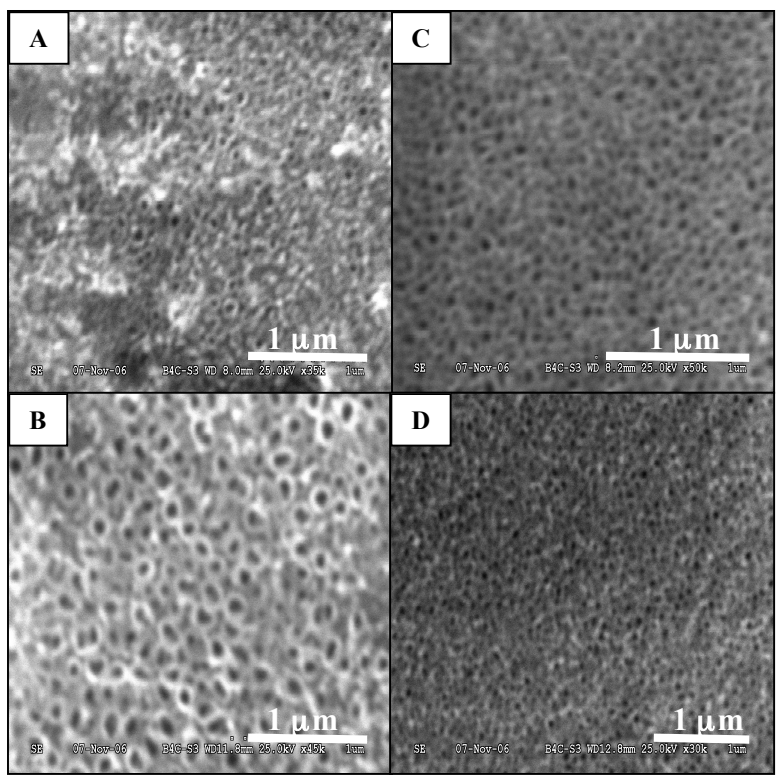

**Figure 1:** SEM images of TiO$_2$ nanopores prepared by anodization for 5 minutes in (A) 0.5 %, (B) 1%, (C) 5%, and (D) 10% HF solutions.

The SEM images (figure1) show the TiO$_2$ nanopores structures which were developed by the application of constant potential for 5 minutes. Keeping the anodizing time constant, the concentration of HF was altered to obtain the nanostructures. With a progressive increase in HF concentration (from 0.5% to 10%), the diameter of the nanopores gradually decreased, approximately from 100 nm diameter to 50 nm. The morphology of the nanostructures underwent a detectable transformation with the change in HF concentration. At low concentrations of HF (figure 1A and 1B), the nanopores exhibited a distinct tube structure. The nanostructures formed at higher HF concentrations (figure 1C and 1D) displayed pores with interconnected walls. The results show that the nanopore dimensions can be modified by controlling HF concentration.

To examine the effect of the length of anodization time on the formation of nanopores, the anodic oxidation of titanium was carried out at 0.5 % HF concentration with reaction times from 5 min to 30 minutes (figure 2).

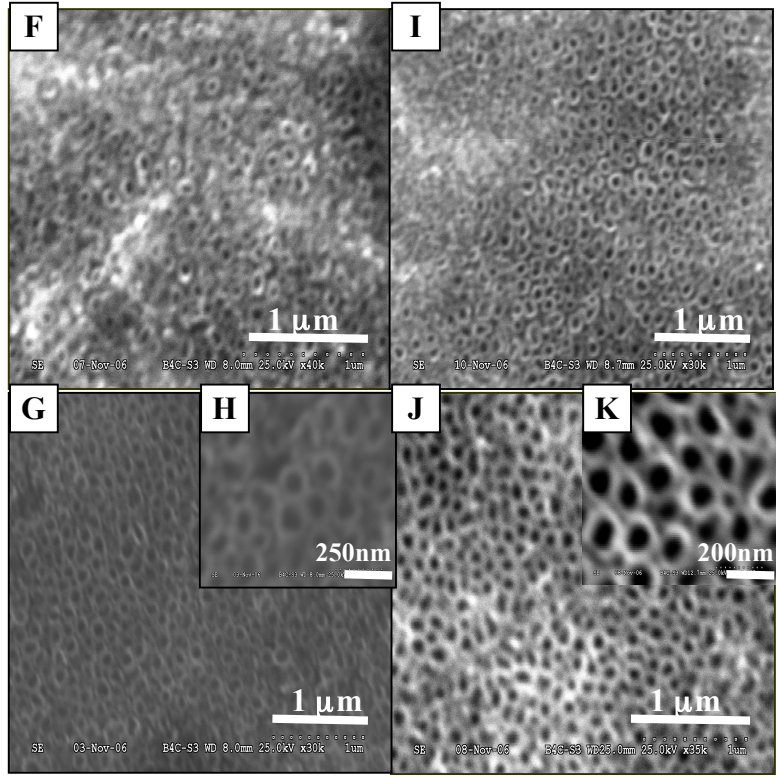

**Figure 2:** TiO$_2$ nanopores prepared by anodization in 0.5%HF for (F) 5 mins, (G) 10 mins, (I) 20 mins, and (J) 30 minutes. Inserts H and K show the structures formed at 10 and 30 minutes.

The pore size remained largely unaffected with the increase of anodization time. Nanopores formed at different anodization times displayed diverse morphologies. As anodization time increased, the nanopores showed a noticeable transformation from tube like structures to pore networks (figure 2). An enhancement of the uniformity of the pore distribution was observed at longer anodization times (figure 2, I and J).

A similar investigation was carried out at higher concentration of HF (figure 3). Titanium was anodized in 5% HF solution with varying reaction times from 5 minutes to 20 minutes.

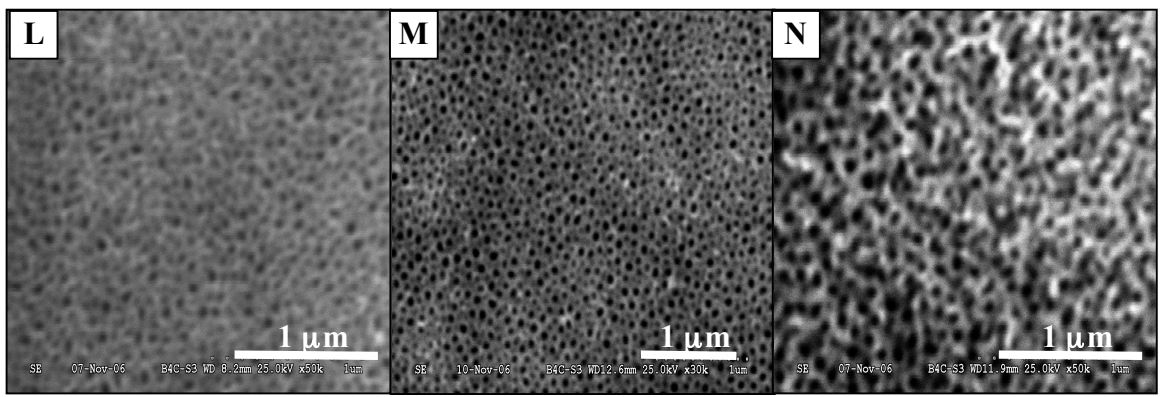

**Figure 3:** TiO$_2$ nanopores prepared by anodization in 5% HF for (L) 5 mins, (M) 10 mins, and (N) 20 minutes.

Once again, varying the anodization time effected morphological changes in the nanopores. Long anodization times resulted in merging nanopore structures (figure 3 N), whereas, at shorter anodization times, distinct nanopores were obtained (figure 3, L and M).

The formation of a titanium oxide nanoporous layer on the metal foil is controlled by three processes, field-enhanced oxidation of titanium foil, field–assisted oxide dissolution, and chemical oxide dissolution [13]. During field–enhanced oxidation, titanium metal is converted to its oxide form, as oxygen-containing ions in the solution move towards the interface of the metal and its oxide. During field-assisted oxide dissolution, titanium ions migrate from the metal-oxide interface through the oxide layer and dissolve into the solution. HF accelerates the rate of dissolution by consuming titanium ions in solution [14]. Since this mechanism includes no restriction to pore widening, the extensive dissolution which occurs over long anodization times leads to pore merging and the eventual loss of the nanopore structures.

**CONCLUSIONS**

This study qualitatively describes the dependence of the structure and morphology of TiO$_2$ nanopores on the two primary parameters of the anodic oxidation process at constant voltage. The pore size of TiO$_2$ nanostructures is affected by the electrolyte concentration during anodization, whereas the morphology and distribution of the nanopores is affected by both electrolyte concentration and anodization time. The study found that TiO$_2$ nanostructures show a unique tube to pore transition with an increase in anodization time and electrolyte concentration at constant voltage. The knowledge acquired in this study can be applied to modulate the growth of nanostructured oxide materials, for use in nanopore based biosensing devices.

## ACKNOWLEDGMENTS

The authors thank Professor Harry Gafney and Mr. François LaForge, of the Department of Chemistry at Queens College, and Dr. Baohe Chang for their assistance. The work was supported in part by a grant from The City University of New York PSC-CUNY Research Award program and in part by the James D. Watson NYSTAR Award.